\documentclass[onecolumn,10pt,showpacs]{revtex4}
\usepackage{graphicx}
\usepackage{epsfig}
\usepackage{bm}
\usepackage{amsfonts}

\input epsf

\begin{document}

\title{\Large Dynamical Study of DBI-essence in Loop Quantum
\\ Cosmology and Braneworld  }

\author{\bf Jhumpa Bhadra$^1$\footnote{bhadra.jhumpa@gmail.com}
and Ujjal Debnath$^1$\footnote{ujjaldebnath@yahoo.com ,
ujjal@iucaa.ernet.in}}

\affiliation{$^1$ Department of Mathematics,
Bengal Engineering and Science University, Shibpur, Howrah-711
103, India.\\}

\date{\today}

\begin{abstract}
We have studied homogeneous isotropic FRW model having dynamical
dark energy DBI-essence with scalar field. The existence of
cosmological scaling solutions restricts the Lagrangian of the
scalar field $\phi$. Choosing $p=X g(X e^{\lambda \phi})$, where
$X=-g^{\mu\nu} \partial_\mu \phi \partial_\nu \phi /2$ with $g$ is
any function of $X e^{\lambda \phi}$ and defining some suitable
transformations, we have constructed the dynamical system in
different gravity: (i) Loop Quantum Cosmology (LQC), (ii) DGP
BraneWorld and (iii) RS-II Brane World. We have investigated the
stability of this dynamical system around the critical point for
three gravity models and investigated the scalar field dominated
attractor solution in support of accelerated universe. The role of
physical parameters have also been shown graphically during
accelerating phase of the universe.
\end{abstract}


\maketitle

\sloppy \tableofcontents

\newpage

\section{Introduction}

Cosmic acceleration is on of the most challenging observation of
the cosmology \cite{{Riess},{Perlmutter2}}. The reason for this
accelerating universe is termed as {\it dark energy} which
dominates the universe (70\% of the universe) having large
negative pressure and violates the strong energy condition
\cite{{Sahni1}, {Peebles},{Padmanabhan}}. The most popular
candidate of dark energy is the cosmological constant $\Lambda$
\cite{Sami} whose EoS parameter is $w=-1$. This kind of dark
energy shows that the universe will accelerate forever. Another
kind of dark energy dubbed as quintessence ($w<-1/3$) explains
that the acceleration will replaced by deceleration in far future.
And for phantom energy ($w<-1$), acceleration will change to super
acceleration which  will eventually destroy every stable
gravitational structure \cite{Zhang}. There are also other
candidates of dark energy like Chaplygin gas \cite{Kamenshchik},
modified Chaplygin gas (MCG) \cite{Banerjee}, Tachyonic
field \cite{Sen}, DBI-essence \cite{Martin}, K-essence \cite{Armendariz-Picon} and so on.\\

There are numerous works done on dark energy on the theory of
Einstein's classical general relativity (GR). But, most physicists
deemed that the gravity should be quantized. Loop quantum gravity
(LQG) is an outstanding effort to describe the quantum effect of
our universe. In this theory classical space time continuum is
replaced by discrete quantum geometry. Now a days several
cosmological (interacting dark energy model) models are studied in
the frame work of LQC. Wu and Zhang \cite{Wu} studied the
cosmological evolution in LQC for the quintessence model. Chen et
al \cite{Chen} provided the parameter space for the existence of
the accelerated scaling attractor in LQC with more general
interacting term. When the Modified Chaplying Gas coupled to dark
matter in the universe is described in the frame work LQC by
Debnath et al \cite{jamil} who resolved the famous
cosmic coincidence problem in modern cosmology.\\

There is another modification on gravity (Brane-gravity) which
also exhibits the acceleration of the present day universe. It is
proposed that our universe is a 3-brane embedded in a four
dimensional space. An important ingredient of the brane world
scenario is that the standard matter particles and forces are
confined on the 3-brane and the only communication between the
brane and bulk is through gravitational interaction (i.e., gravity
can freely propagate in all dimensions) or some other dilatonic
matter. In the review \cite{Rubakov,Maartens,Brax,Csa} there is
different applications with special attention to cosmology in
Brane-gravity. In this work
we consider the two most popular brane models, namely DGP and RS II branes.\\

Regarding cosmological acceleration,  dark energy with  energy
density of scalar field act subdominant  during radiation and dark
matter eras and acts dominant  at late times. Dynamical system
theory has been applied with great success in cosmology and
astrophysics within the context of general relativity. This theory
are used to describe the behaviour of complex dynamical systems
usually by constructing differential equations. This theory deals
with a long term qualitative behaviour of the formed first order
differential equations. It does not concentrate to find the
precise solutions of the system but provide answers like whether
the system is stable for long time and whether the stability
depends on the initial conditions. Besides the other scientific
fields this theory is now become widely useful in the research of
cosmology. In the construction  of  different dark energy model
cosmological scaling solutions work significant role
\cite{{Copeland},{Liddle},{Macorra},{Tsujikawa}}. Tsujikawa et al
\cite{{ Tsujikawa},{ Piazza}} proved that scaling solution exists
for coupled dark energy whenever they restrict to the form of the
field Lagrangian $p(X,\phi)=X g(X e^{\lambda \phi})$ where
$X=-g^{\mu\nu} \partial_\mu \phi \partial_\nu \phi /2$ and $g$ is
any function of $ X e^{\lambda \phi}$. In reference \cite{Naskar},
they also considered the interacting model in these Lagrangian
form and studied the stability of fixed points for several
different dark energy models for ordinary (phantom) field,
dilatonic ghost condensate and (phantom) tachyon. Our main aim of
this work is to examine the nature of the different physical
parameters for the universe around the stable critical points in
LQC and two brane world models (DGP and RS II) in presence of
DBI-esssence type dark energy along with dark matter with suitable
interaction term. With the evolution of the universe we find the
effective state parameter $w_{eff}$, Critical densities for dark
energy ($\Omega_\phi$) and for dark matter ($\Omega_m$) and
examine future dominance nature of kinetic energy
and potential energy. \\

In this work, we have considered the field Lagrangian
$p(X,\phi)=X g(X e^{\lambda \phi})$ and studied the dark energy
model in different gravity theories like (i) LQC, (ii) DGP
Brane-world, (iii) RS II Brane-world. We also derive the critical
point of the dynamical system in different gravity and analyze the
stability. Also we do the numerical simulation for LQC,
DGP-Brane-world and RS II Brane-world models. Some fruitful
conclusions are drawn in
section V.\\

\section{Basic Equations in DBI-essence}

The action of the Dirac-Born-Infeld (DBI) scalar field $\phi$ can
be written as (choosing $8\pi G = c = 1$) \cite{Yamaguchi}

\begin{eqnarray}
S_{DBI}=-\int d^4 x
\sqrt{-g}\left[T(\phi)\sqrt{1-\frac{\dot{\phi}^2}{T(\phi)}}-T(\phi)+V(\phi)\right]
\end{eqnarray}

where $V (\phi)$ is the self-interacting potential and $T (\phi)$
is the warped brane tension. The kinetic term of the above action
is non-canonical. Physically, this originates from the fact that
the action of the system is proportional to the volume traced out
by the brane during its motion. This volume is given by the
square-root of the induced metric which automatically leads to a
DBI kinetic term.\\

From the above action, it is easy to determine the energy density
and pressure of the DBI-essence scalar field which are
respectively given by
\begin{eqnarray}
\rho_{\phi}=(\gamma -1)T(\phi)+V(\phi)
\end{eqnarray}
and
\begin{eqnarray}
p_{\phi}=\frac{(\gamma -1)}{\gamma}T(\phi)-V(\phi)
\end{eqnarray}
where $\gamma$ is given by
\begin{eqnarray}
\gamma=\frac{1}{\sqrt{1-\frac{\dot{\phi}^2}{T(\phi)}}}
\end{eqnarray}

From above expression, we observe that $T(\phi)>\dot{\phi}^2$ and
thus $\gamma>1$. We consider a spatially flat Friedmann-Lemaitre-
Robertson-Walker (FLRW) Universe containing a perfect fluid and a
scalar field $\phi$. Assuming that there is an interaction between
scalar field (dark energy) and the perfect fluid (dark matter), so
they are not separately conserved. The energy balance equations
for the interacting dark energy and dark matter can be expressed
as \cite{Piazza}

\begin{eqnarray}
\dot{\rho}_{\phi}+3H(1+w_{\phi})\rho_{\phi}=-Q \rho_m \dot{\phi}
\end{eqnarray}
and
\begin{eqnarray}
\dot{\rho}_{m}+3H(1+w_{m})\rho_{m}=Q \rho_m \dot{\phi}
\end{eqnarray}

where $\rho_{m}$ is the energy density of the dark matter, $w_{m}$
is the EoS parameter for the dark matter, $H=\frac{\dot{a}}{a}$
Hubble parameter, $a$ is the scale factor and
$Q>0$ is the coupling between dark energy (DBI-essence) and the dark matter.\\

We define the fractional density of dark energy and dark matter,
$\Omega_{\phi}=\frac{\rho_{\phi}}{3H^2}$ and $\Omega_{m}=\frac{\rho_{m}}{3H^2}$~.\\

To get stable attractor solution we must have $\gamma=$ constant,
$p_\phi=X g(Y)$, is the scalar field pressure density where $Y= X
e^{\lambda \phi}$, $X=-g^{\mu\nu} \partial_\mu \phi
\partial_\nu \phi /2$ with $g$ is any function of $Y$ \cite{Tsujikawa, Piazza}. For DBI-essence
we choose

\begin{eqnarray}
T(\phi)=\frac{\gamma^2}{\gamma^2-1} \dot{\phi}^2 ,~~~~~~~~~V(\phi)=V_0 e^{\lambda \phi}
\end{eqnarray}

Then the pressure $p_\phi$ and the energy density $\rho_\phi$ can
be written in the form

\begin{eqnarray}
p_\phi=X g(Y)\nonumber\\
\rho_\phi= 2X\frac{\partial p_\phi}{\partial X}-p_\phi= X[g(Y)+2Y g'(Y)]
\end{eqnarray}

whenever we choose
\begin{eqnarray}
g(Y)=\frac{2 \gamma^2}{\gamma^2-1}-\frac{V_0}{Y}
\end{eqnarray}
where $'$ denotes the derivative with respect to $Y$. The total cosmic energy density
$\rho=\rho_{\phi}+\rho_{m}$ satisfies the conservation equation $\dot{\rho}+3H(\rho+p)=0$, where $p=p_{\phi}+p_{m}$. \\

\section{Evaluation of dynamical system in LQC}

The modified Friedmann equation for LQC is given by  \cite{{Wu},{Chen},{Fu}}.
\begin{eqnarray}
H^2=\frac{\rho}{3} \left(1-\frac{\rho}{\rho_c}\right)
\end{eqnarray}

Here $\rho_c\equiv \sqrt{3} \pi^2 \eta^3 G^2 \hbar$ is the critical loop quantum density
and $\eta$ is the dimensionless Barbero-Immirzi parameter. It should be noted that
for our LQC model, $\rho<\rho_{c}$~.\\

Consequently we obtain the modified Raychaudhuri equation (using
the conservation law)
\begin{eqnarray}
\dot{H}=-\frac{1}{2}\left(p + \rho \right) (1-2\frac{\rho}{\rho_c})
\end{eqnarray}

We introduce the following dimensionless quantities

\begin{eqnarray}
x=\frac{\dot{\phi}}{\sqrt{6} H},~~~~~~~~ y=\frac{e^{-\lambda \phi/2}}{\sqrt{3} H},~~~~~~~~~ z=\frac{\rho}{\rho_c}
\end{eqnarray}

We see that $y$ and $z(<1)$ must be non-negative, but $x$ may or
may not be positive depends on the nature of $\dot{\phi}$.
Substituting the expressions of $x,~y$ and $z$ in equations (5),
(6) and (11), we obtain the first order differential equations in
the form of autonomous system as follows:
\begin{eqnarray}
\frac{d x}{d N}=-3x+\frac{3 x}{2}\left[A(1-w_m)x^2-\frac{(1+\omega_m)\left\{1+V_0 y^2(z-1)\right\}}{z-1}\right](1-2z)+\nonumber\\
\frac{\sqrt{6} Qx^2\left\{1+Ax^2(z-1)+V_0 y^2(z-1)\right\}}{2 Ax^2(z-1)}+\frac{\sqrt{6} \lambda V_0 y^2 (3x^2-4 y^3)(1-z)}{2 Ax^2(z-1)}\\
\frac{dy}{dN}=\frac{3 y}{2}\left[A(1-w_m)x^2-\frac{(1+w_m)\left\{1+V_0 y^2(z-1)\right\}}{z-1}\right](1-2z)-\frac{\sqrt{6}}{2}\lambda xy\\
\frac{dz}{dN}=-3 \left[A(1-w_m)x^2-\frac{(1+w_m)\left\{1+V_0 y^2(z-1)\right\}}{z-1}\right](1-z)z\\
\frac{1}{H}\frac{dH}{dN}=-\frac{3}{2}\left[A(1-w_m)x^2-\frac{(1+w_m)\left\{1+V_0 y^2(z-1)\right\}}{z-1}\right](1-2z)
\end{eqnarray}

where $A=\frac{2 \gamma^2}{\gamma^2-1}$, $N=\ln a$ being the
number of $e$-folds and $a$ being the scale factor. In terms of
the new variables $x,~y,~z$ we obtain the following physical
parameters

\begin{eqnarray}
\Omega_\phi=Ax^2+V_0 y^2,~~~~~~~~~~~~w_\phi=\frac{Ax^2-V_0y^2}{Ax^2+V_0 y^2}\\
w_{eff}=\frac{p_\phi+p_m}{\rho_\phi+\rho_m}=\frac{w_m+(Ax^2-V_0 y^2)(1-z)}{1+Ax^2(1-z)+V_0y^2(1-z)}
\end{eqnarray}

The fraction density function as $\Omega_\phi=\frac{\rho_\phi}{3 H^2}$ and $\Omega_m=\frac{\rho_m}{3 H^2}$ satisfying
\begin{eqnarray}
\Omega_\phi+\Omega_m+\Omega_{LQC}=1\nonumber
\end{eqnarray}

where, $\Omega_{LQC}=\frac{\rho}{\rho-\rho_c}$ is the density
parameter due to the effect of LQC. Since $\rho<\rho_{c}$, so $\Omega_{LQC}<0$ in our case.\\

The new variables ($x,y,z$) have been drawn in figure 1 with
respect to $N=\ln a$ and seen that all are positive oriented due
to the expansion of the universe. Also $\Omega_{\phi}$,
$\Omega_{m}$ and $w_{eff}$ have been drawn in figure 2.
$\Omega_{m}$ shows the lower value ($<1$) and $\Omega_{\phi}$
shows the value $>1$ in evolution, so $\Omega_{\phi}$ gets higher
value than $\Omega_{m}$. So in late stage, the dark energy
(DBI-essence) dominates over dark matter. Also $w_{eff}$ gives the
negative value less than $-0.5$ which shows the dark energy
dominated phase of the universe.

\begin{figure}
\epsfxsize = 2.3 in \epsfysize = 1.7 in
\epsfbox{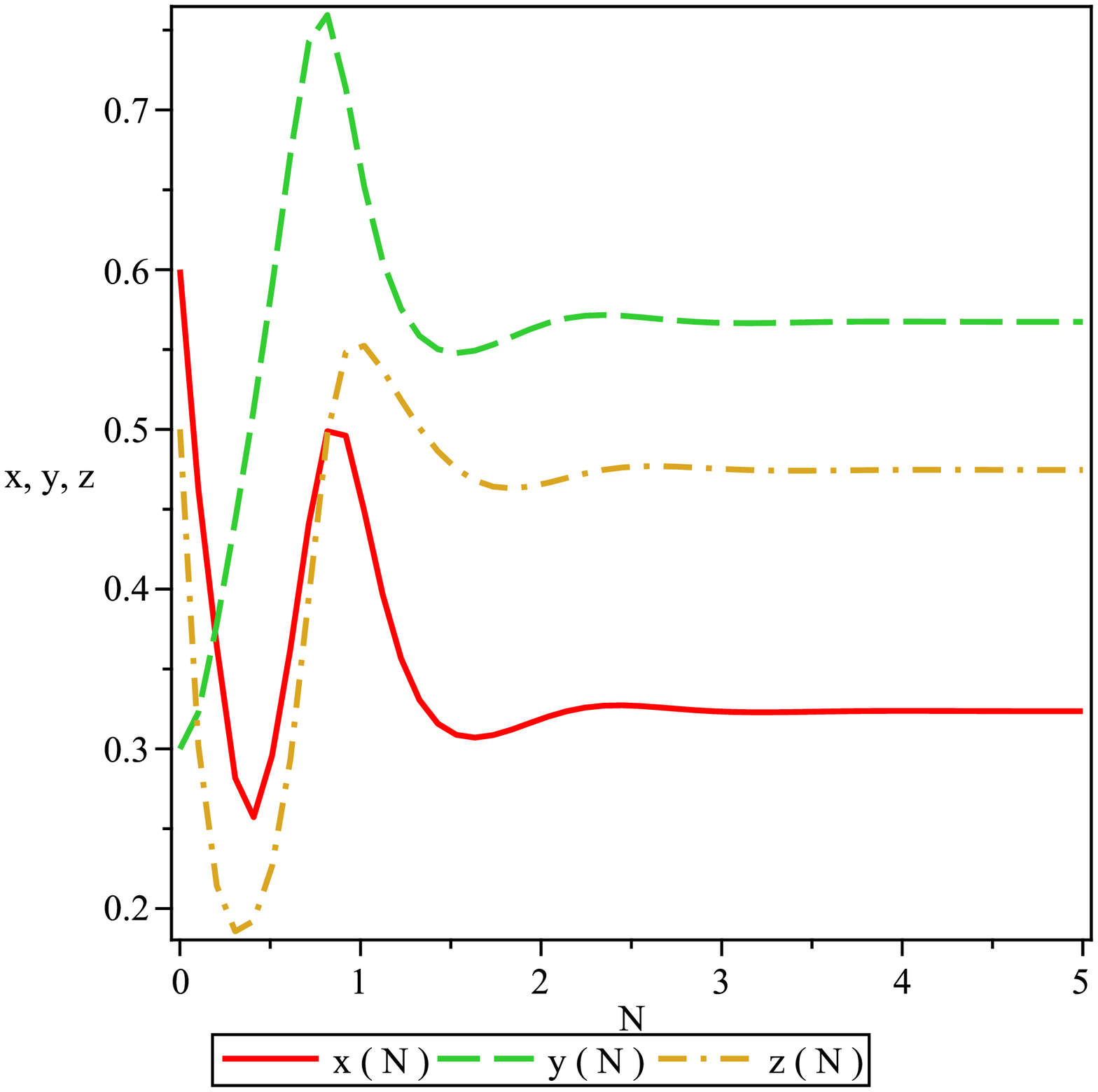}~~~\epsfxsize = 2.3 in \epsfysize =1.7 in
\epsfbox{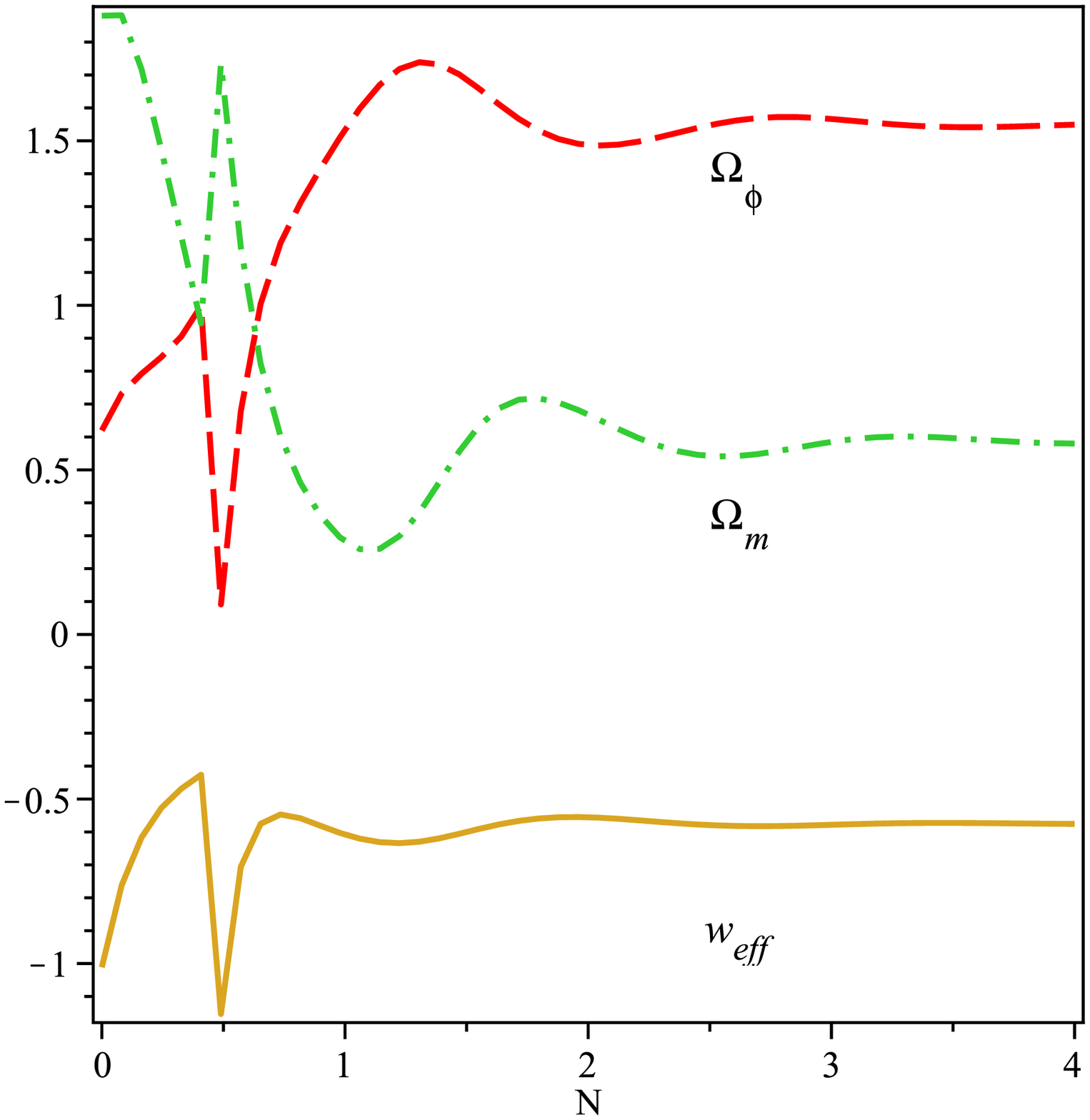}\\
~~FIG.1~~~~~~~~~~~~~~~~~~~~~~~~~~~~~~~~~~~~~~~~~~~~~~~~~~~~~~~~~~~~~~~~~~~~FIG.2\\
\caption{Evaluation of $x$, $y$, $z$ with respect to $N$ in LQC
model taking $\gamma=1.2$, $Q=0.05$, $V_0=8$, $\lambda=0.5$ and
$w_m=0.01$.} \caption{Evaluation of $\Omega_\phi$, $\Omega_m$ and
$w_{eff}$ with respect to $N$ in LQC model taking $\gamma=0.8$,
$Q=0.05$, $V_0=7.6$, $\lambda=0.5$ and $w_m=0.01$.}
\end{figure}

\subsubsection{Critical points:}

The critical points can be obtained by setting $\frac{dx}{dN}=0$,
$\frac{dy}{dN}=0$ and $\frac{dz}{dN}=0$ and are presented in the
following table.\\

{\bf Table 1:} The critical points ($x_{c},~y_{c},~z_{c}$) and the
corresponding values of the density parameter $\Omega_{\phi}$.

\begin{center}
\begin{tabular}{|l|}
\hline\hline
~~No.~~$x_c$~~~~~~~~~~~~$y_c$~~~~~~~~~~~~~~$z_c$~~~~~~~~~~~~~~~~~~~~~~~~$\Omega_\phi$\\ \hline
\\\\
~~(i)~~$\frac{1}{\sqrt{A}}$~~~~~~~~~~~~~0~~~~~~~~~~~~~~~0~~~~~~~~~~~~~~~~~~~~~~~~~~1\\\\
~~(ii)~$-\frac{1}{\sqrt{A}}$~~~~~~~~~~~0~~~~~~~~~~~~~~~0~~~~~~~~~~~~~~~~~~~~~~~~~~1\\\\
~~(iii)~$\frac{\sqrt{\frac{2}{3}}Q}{A(w_m-1)}$~~~~~~0~~~~~~~~~~~~~~~0~~~~~~~~~~~~~~~~~~~~~$\frac{2 Q^2}{3A(1-w_m)^2}$\\\\
~~(iv)~$\frac{\sqrt{6}(1+w_m)}{2 Q}$~~~~~0~~~~~~~~$\frac{3A(1-w_m^2+2Q^2)}{3A(1-w_m^2)}$~~~~~~~~~~~~$\frac{3A(1+w_m)^2}{2 Q^2}$\\
\\\hline\hline
\end{tabular}
\end{center}

%

From the Table 1, we see that the components of $y_{c}$ is equal
to zero for above four critical points. The value of
$\Omega_{\phi}=1$ for the critical points given in (i) and (ii).
These provide the accelerated phase of the universe. Similar
nature happen for other two critical points (iii) and (iv), but
these depend on the interaction term $Q$, $A$ and $w_{m}$.

\subsubsection{Stability of the model:}

Now the stability around the critical points can by determined by
the sign of the corresponding eigen values. If the eigen values
corresponding to the critical point are all negative, the critical
points are stable node, otherwise unstable. The eigen values for
the above critical points are obtained as in the following:\\

{\bf Table 2:} The eigen values corresponding to the critical
points ($x_{c},~y_{c},~z_{c}$).

\begin{center}
\begin{tabular}{|l|}
\hline\hline
~~No:~~~~~~~~Value1~~~~~~~~~~~~~~~~~~~~~~~~~~~~~~~~~~~~~Value2~~~~~~~~~~~~~~~~~~~~~~~~~~~~~~~~~~~Value3
\\ \hline
\\\\
~~(i)~~~~~~~~~~-6~~~~~~~~~~~~~~~~~~~~~~~~~~~~~~~~~~~$3+\frac{\sqrt{6}Q}{\sqrt{A}}-3w_m$~~~~~~~~~~~~~~~~~~~~~~~~~~~~$3-\sqrt{\frac{3\lambda}{2A}}$
\\\\
~~(ii)~~~~~~~~~-6~~~~~~~~~~~~~~~~~~~~~~~~~~~~~~~~~~~$3-\frac{\sqrt{6}Q}{\sqrt{A}}-3w_m$~~~~~~~~~~~~~~~~~~~~~~~~~~~~$3+\sqrt{\frac{3\lambda}{2A}}$
\\\\
~~(iii)~~~$-\frac{3}{2}+\frac{3w_m}{2}+\frac{Q^2}{A(1-w_m)}$~~~~~~~~~~~$-\frac{2Q^2}{A(1-w_m)}-3(1+w_m)$~~~~~~~~~~~~~~$-\frac{3A(1-w_m^2)+2Q(Q+\lambda)} {2A(-1+w_m)}$\\\\
~~(iv)~~~~~~$-\frac{3}{2}\left(1+R\right)$~~~~~~~~~~~~~~~~~~~~~~~~~~$\frac{3}{2}\left(-1+R\right)$~~~~~~~~~~~~~~~~~~~~~~~~~~~~~~~$\frac{3(1+w_m)\lambda}{2 Q}$\\\\
~~~~~~~~~~~~~~~~~~~~~~~~~~~~~~~~~~~~~~~~~~~where, $R=\sqrt{\frac{6A(-1+w_m)(1+w_m)^2-Q^2(3+4w_m)}{Q^2}}$

\\\hline\hline
\end{tabular}
\end{center}

\vspace{.2in}

From the Table 2, we observe in the following:\\

(a) One eigen value for the critical point (i) is positive, since
$3 (1-w_m)+\frac{\sqrt{6}Q}{\sqrt{A}}>0$, so around this critical
point system is not stable.\\

(b) One eigen value for the critical point (ii) is positive,
since, $3+\sqrt{\frac{3\lambda}{2A}}>0$, so around
this critical point system is not stable.\\

(c) If $\gamma^2<1$ and $Q(Q+\lambda)>\frac{3 A (1-w_m^2)}{2}$,
the eigen values of the critical point (iii) are all negative and hence the system is stable (node).\\

(d) All eigen values of the critical point (iv) are negative if
$\lambda<0$ and $R<1$, so the system may be stable otherwise the
system will be unstable.\\

\section{Brane World}

\subsection{Basic equations in DGP Brane model}

An effectual model of brane-gravity is the Dvali-Gabadadze-Porrati
(DGP) braneworld model \cite{{Dvali},{Deffayet}} that represents
our 4-dimensional universe to a FRW brane embedded in a
5-dimensional Minkowski bulk. It explains the origin of DE as the
gravity on the brane escaping to the bulk at large scale. On the
4-dimensional brane the action of gravity is proportional to
$M_p^2$. That action is proportional to the corresponding quantity
in 5-dimensions in the bulk. The modified Friedmann equation in
DGP brane model considering flat, homogeneous and isotropic brane
is given by

\begin{eqnarray}
H^2=\left(\sqrt{\frac{\rho}{3}+\frac{1}{4 r_c^2}}+\epsilon \frac{1}{2 r_c}\right)^2
\end{eqnarray}

where $\rho$ is the cosmic fluid energy density,
$H=\frac{\dot{a}}{a}$, Hubble parameter and $r_c=\frac{M_p^2}{2
M_5^2}$ is the crossover scale which resolve the transition from
4D to 5D behavior and $\epsilon=\pm1$. Corresponding to
$\epsilon=+1$ the we have standard DGP$(+)$ model which is self
accelerating model without any form of DE, and effective $w$ is
always non-phantom. However for $\epsilon = -1$, we have DGP$(-)$
model which does not self accelerate but requires DE on the brane.
Using (18), the modified Raychaudhuri equation becomes (choosing
$8\pi G=c=1$)

\begin{eqnarray}
\left(2H-\epsilon \frac{1}{r_c}\right)\dot{H}=-H(\rho+p)
\end{eqnarray}

\subsubsection{Dynamical system }

To get dynamical analysis of our DGP brane world model of the
universe, we define the following dimensionless quantity

\begin{eqnarray}
x=\frac{\dot{\phi}}{\sqrt{6}H},  ~~~~y=\frac{e^{-\lambda \phi/2}}{\sqrt{3}H} ~~~~~z=2H-\frac{\epsilon}{r_c}
\end{eqnarray}

We see that $y$ must be non-negative, but $x$ and $z$ may or may
not be positive. When $\phi$ is increasing, $x$ must be positive
and $\phi$ decreases implies $x$ is negative. Also $\epsilon=-1$
implies $z>0$, but for $\epsilon=+1$, the value of $z$ may or may
not be positive. Now we introduce the fraction density parameters
as $\Omega_\phi=\frac{\rho_\phi}{3 H^2}$ and
$\Omega_m=\frac{\rho_m}{3 H^2}$ satisfying
\begin{eqnarray}
\Omega_\phi+\Omega_m+\Omega_{DGP}=1
\end{eqnarray}

where, $\Omega_{DGP}=\frac{\epsilon}{r_c H}$ is the density
parameter due to the effect of DGP brane world with the physical
parameters
\begin{eqnarray}
\Omega_\phi=Ax^2+V_0 y^2,~~~~~w_\phi=\frac{Ax^2-V_0y^2}{Ax^2+V_0 y^2}~~~~~~~~~~~~~~\\
w_{eff}=\frac{p_\phi+p_m}{\rho_\phi+\rho_m}=\frac{4 r_c^2 (Ax^2-V_0 y^2)+ w_m \left[-1-4r_c^2 (-1+Ax^2+V_0 y^2)-4 r_c \epsilon+\epsilon^2\right)] }{(2 r_c+\epsilon)^2-1}
\end{eqnarray}
\\
Using all this and defining $N=\ln a$ (the number of $e$-folds) we
get the system of equations as follows:

\begin{eqnarray}
\frac{dx}{dN}=-3x+\frac{3 x}{2 z}\left(z+\frac{\epsilon}{r_c}\right)\left(2Ax^2-\frac{(1+w_m)\{1+4 r_c^2(-1+A x^2+V_0 y^2)+4r_c\epsilon-\epsilon^2\}} {4 r_c^2}\right)\nonumber\\
+\frac{\sqrt{6} [Q\{1+4 r_c^2(-1+A x^2+V_0 y^2)+4r_c\epsilon-\epsilon^2\}+4 V_0 \lambda r_c^2 y^2]}{8 A r_c^2}
\end{eqnarray}

\begin{eqnarray}
\frac{dy}{dN}=\frac{y}{2}\left[\frac{3}{z} \left(z+\frac{\epsilon}{r_c}\right)\left(2Ax^2-\frac{(1+w_m)\{1+4 r_c^2(-1+A x^2+V_0 y^2)+4r_c\epsilon-\epsilon^2\}} {4 r_c^2}\right)+\sqrt{6}\lambda x\right]
\end{eqnarray}

\begin{eqnarray}
\frac{dz}{dN}=-\frac{3}{2z}\left(z+\frac{\epsilon}{r_c}\right)^2\left(2Ax^2-\frac{(1+w_m)\{1+4 r_c^2(-1+A x^2+V_0 y^2)+4r_c\epsilon-\epsilon^2\}} {4 r_c^2}\right)
\end{eqnarray}

\begin{eqnarray}
\frac{1}{H} \frac{dH}{dN}=-\frac{3}{2 z}\left(z+\frac{\epsilon}{r_c}\right)\left(2Ax^2-\frac{(1+w_m)\{1+4 r_c^2(-1+A x^2+V_0 y^2)+4r_c\epsilon-\epsilon^2\}} {4 r_c^2}\right)
\end{eqnarray}

The new variables ($x,y,z$) has been drawn in figures 3 and 5 with
respect to $N=\ln a$ for DGP($+$) and DGP($-$) models
respectively. In all the cases, $x$ has shown to be negative i.e.,
the DBI scalar field $\phi$ decreases during expansion, but $y$
and $z$ keeps positive sign. The effective EoS parameter $w_{eff}$
and the density parameters $\Omega_{\phi},~\Omega_{m}$ are shown
in figures 4 and 6 for DGP($+$) and DGP($-$) models respectively.
During expansion, the $\Omega_{\phi}$ increases and $\Omega_{m}$
decreases which show the dark energy dominates at late times. Also
$w_{eff}$ decreases from some $-0.4$ to $-1$ for both models,
which also shows the dark energy dominated phase of the universe.\\

\vspace{1in}
\begin{figure}[!h]
\epsfxsize = 2.3 in \epsfysize = 1.7 in
\epsfbox{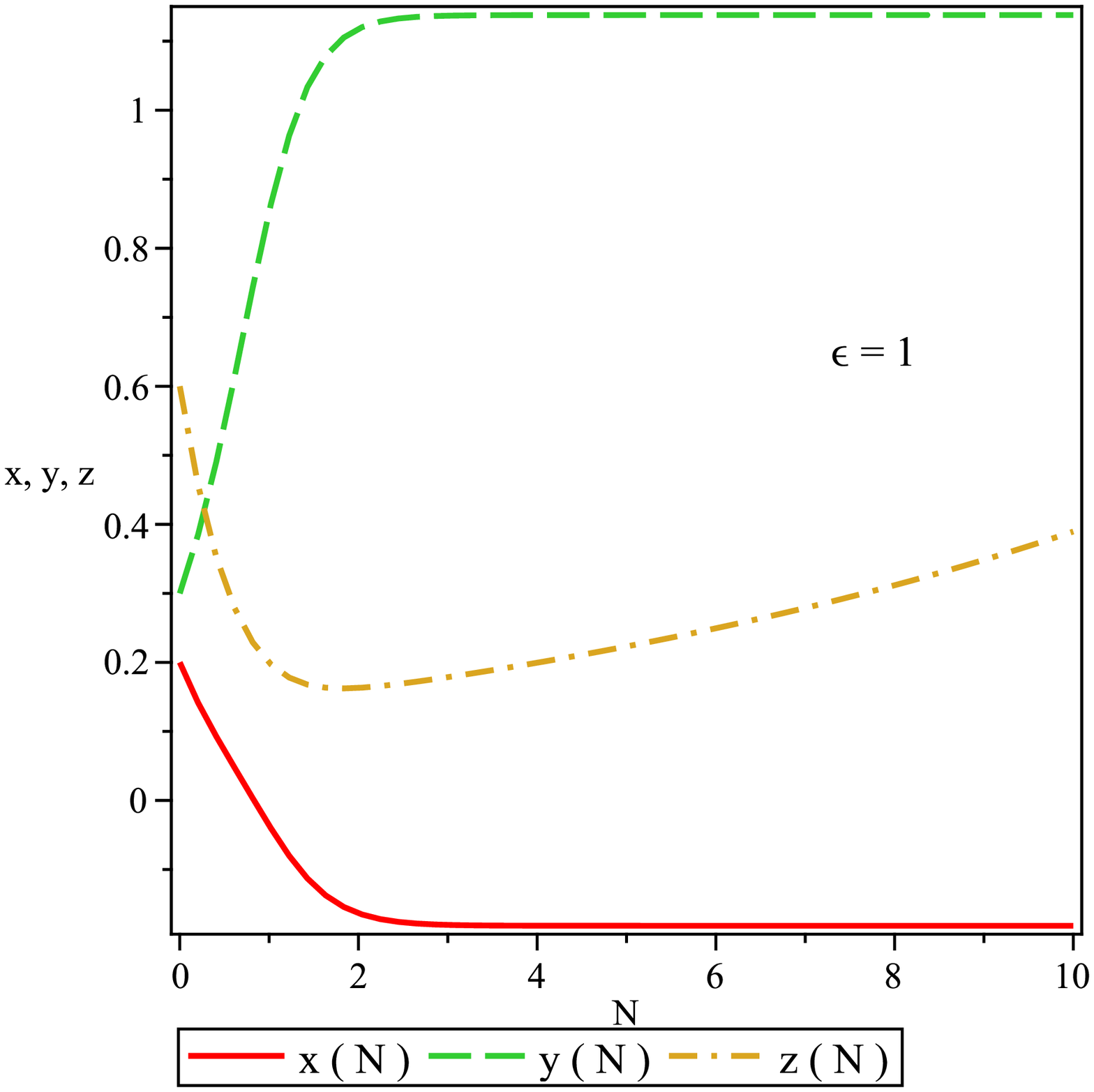}~~~\epsfxsize = 2.3 in \epsfysize =1.7 in
\epsfbox{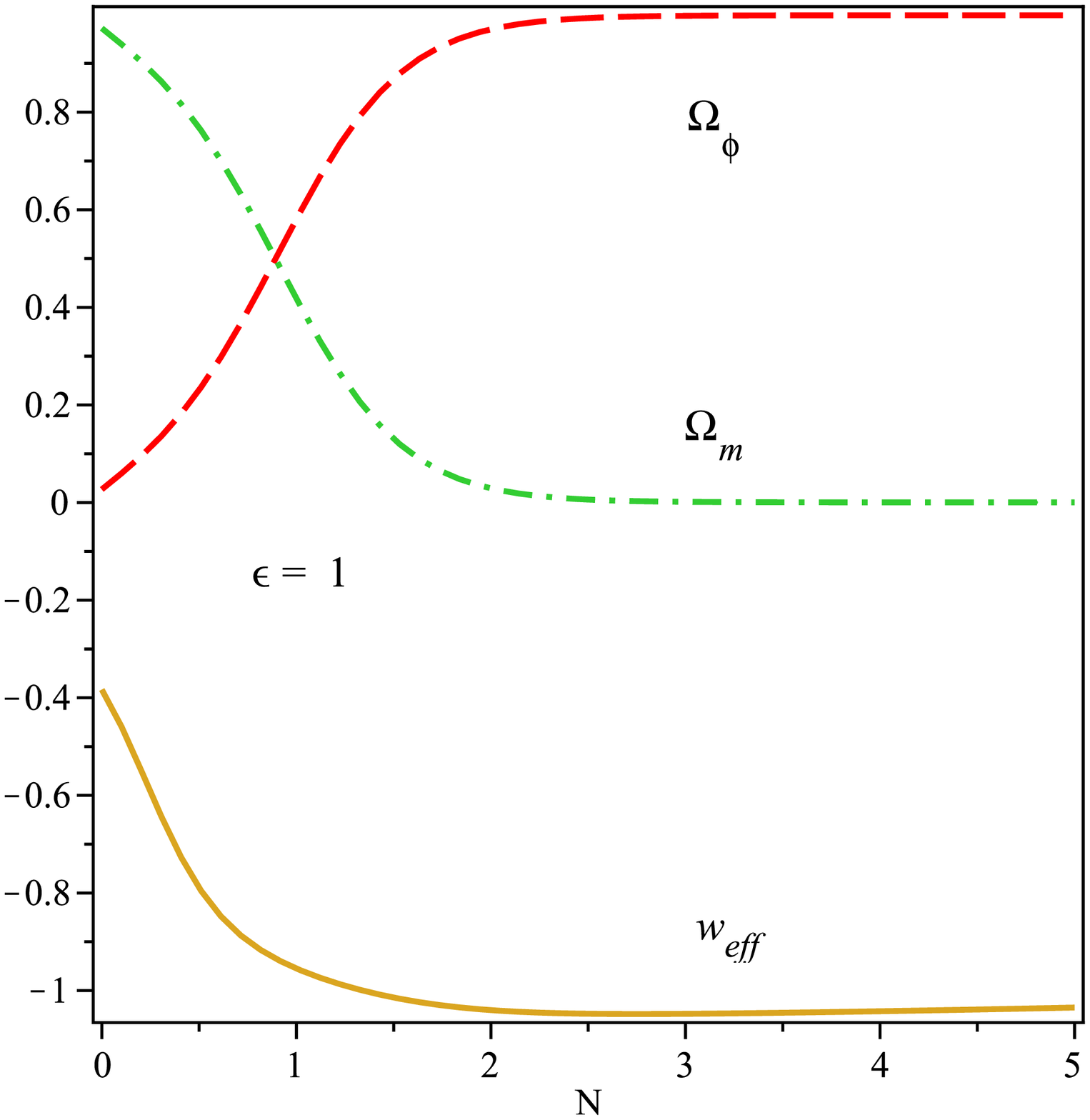}\\
~~FIG.3~~~~~~~~~~~~~~~~~~~~~~~~~~~~~~~~~~~~~~~~~~~~~~~~~~~~~~~~~~~~~~~~~~~~FIG.4\\
\caption{Evaluation of $x$, $y$, $z$ with respect to $N$ in DGP
(+) Brane model ($\epsilon=1$)taking $\gamma=0.6$, $Q=0.05$,
$V_0=0.8$, $\lambda=0.5$, $w_m=0.01$ and $r_c=1000$.}
\caption{Evaluation of $\Omega_\phi$, $\Omega_m$ and $w_{eff}$
with respect to $N$ in DGP (+) Brane model ($\epsilon=1$) taking
$\gamma=0.6$, $Q=0.05$, $V_0=0.8$, $\lambda=0.5$, $w_m=0.01$ and
$r_c=1000$.} \vspace{.2cm}
\end{figure}

\vspace{1in}
\begin{figure}[!h]
\epsfxsize = 2.3 in \epsfysize = 1.7 in
\epsfbox{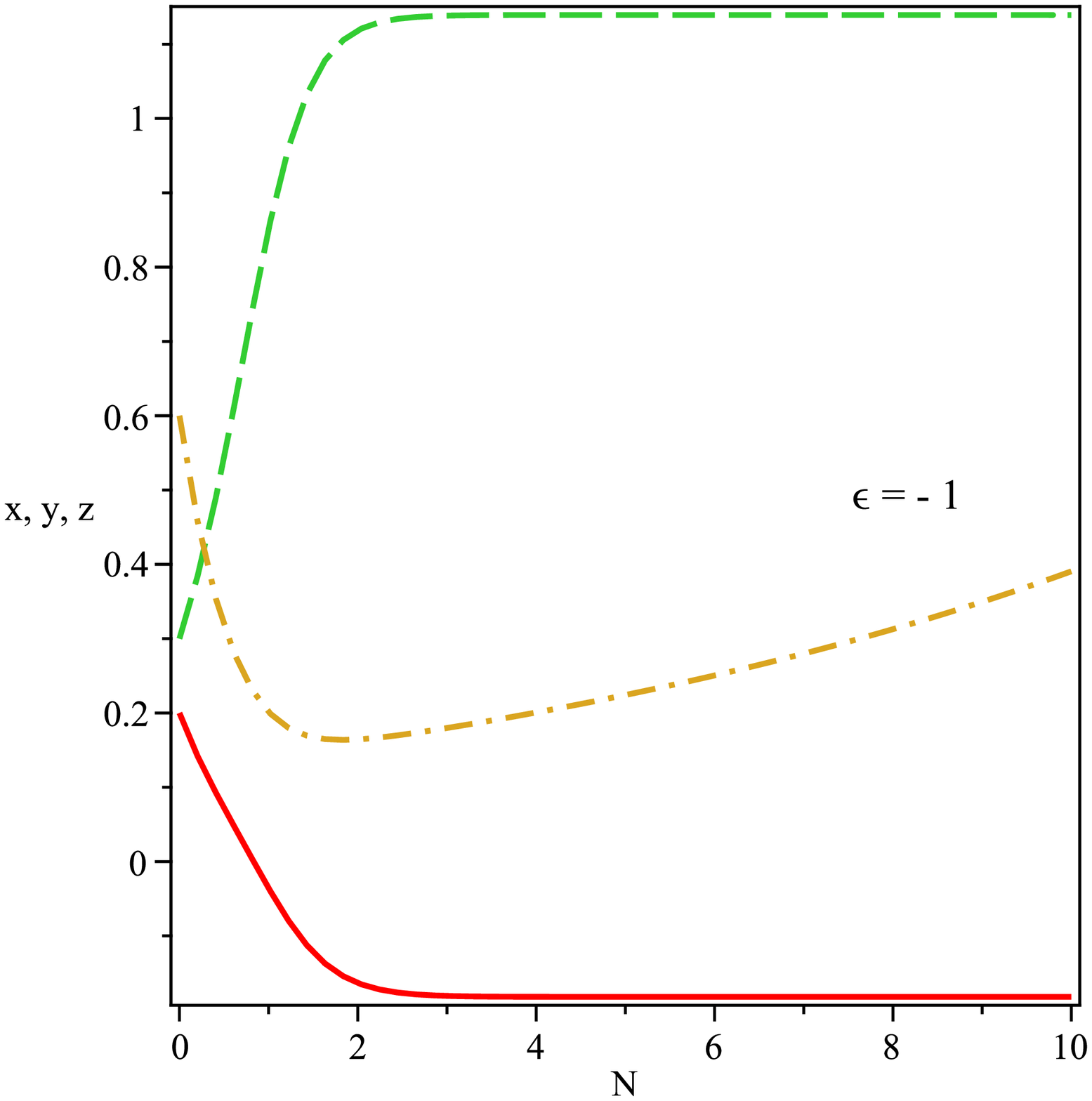}~~~\epsfxsize = 2.3 in \epsfysize =1.7 in
\epsfbox{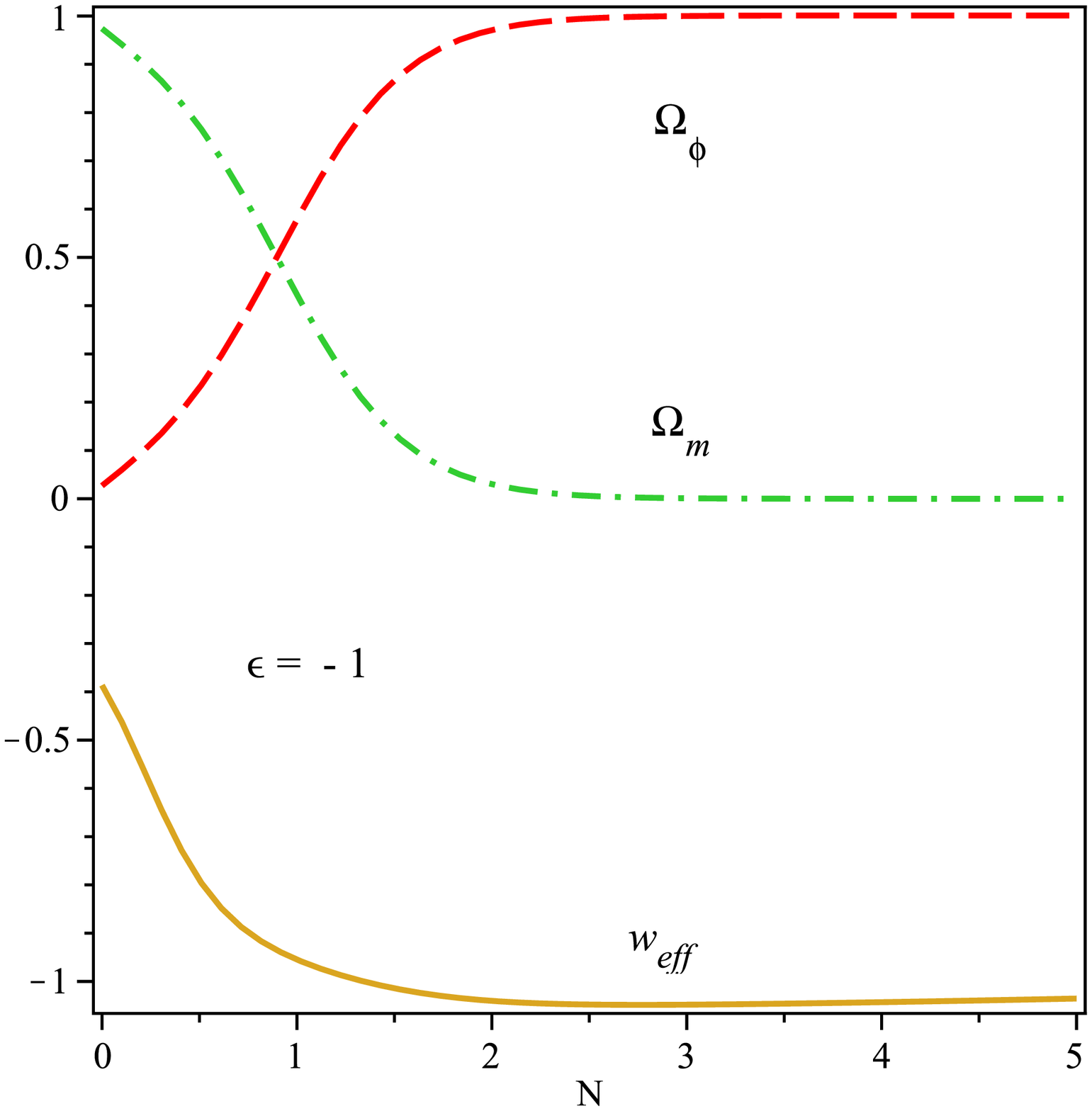}\\
~~FIG.5~~~~~~~~~~~~~~~~~~~~~~~~~~~~~~~~~~~~~~~~~~~~~~~~~~~~~~~~~~~~~~~~~~~~FIG.6\\
\caption{Evaluation of $x$, $y$, $z$ with respect to $N$ in DGP
(-) Brane model ($\epsilon=-1$)taking $\gamma=0.6$, $Q=0.05$,
$V_0=0.8$, $\lambda=0.5$, $w_m=0.01$ and $r_c=1000$.}
\caption{Evaluation of $\Omega_\phi$, $\Omega_m$ and $w_{eff}$
with respect to $N$ in DGP (-) Brane model ($\epsilon=-1$)taking
$\gamma=0.6$, $Q=0.05$, $V_0=0.8$, $\lambda=0.5$, $w_m=0.01$ and
$r_c=1000$.} \vspace{.2cm}
\end{figure}

\subsubsection{Critical Points:}

The critical points can be obtained by setting $\frac{dx}{dN}=0$,
$\frac{dy}{dN}=0$ and $\frac{dz}{dN}=0$. The possible critical
points ($x_{c},y_{c},z_{c}$) and the corresponding values of
$\Omega_{\phi}$ of our DGP model are given by\\
\\
(i) $\left(0,\sqrt{\frac{Q \{-1+(\epsilon-2r_c)^2\}}{4 V_0(Q+\lambda)r_c^2}},-\frac{\epsilon}{r_c}\right)$,  $\Omega_\phi=\frac{Q \{-1+(\epsilon-2r_c)^2\}}{4(Q+\lambda)r_c^2}$\\\\
(ii)$\left(\frac{\sqrt{6} Ar_c^2+\sqrt{Ar_c^2[6Ar_c^2+Q^2\{-1+(\epsilon-2r_c)^2\}]}}{2A Qr_c^2},0,-\frac{\epsilon}{r_c}\right)$,  $\Omega_\phi=\frac{\left(\sqrt{6} Ar_c^2+\sqrt{Ar_c^2[6Ar_c^2+Q^2\{-1+(\epsilon-2r_c)^2\}]}\right)^2}{4A Q^2r_c^4}$\\\\
(iii)$\left(\frac{\sqrt{6} Ar_c^2-\sqrt{Ar_c^2[6Ar_c^2+Q^2\{-1+(\epsilon-2r_c)^2\}]}}{2A Qr_c^2},0,-\frac{\epsilon}{r_c}\right)$, $\Omega_\phi=\frac{\left(-\sqrt{6} Ar_c^2+\sqrt{Ar_c^2[6Ar_c^2+Q^2\{-1+(\epsilon-2r_c)^2\}]}\right)^2}{4A Q^2r_c^4}$\\\\

%

The value of
$\Omega_{\phi}>1$ or $<1$ for the critical points given in (i) to
(iii), depends on the values of the $r_{c}$, $A$ and the
interaction term $Q$.

\subsubsection{Stability of the model:}

Now the stability around the critical points can by determined by
the sign of the corresponding eigen values. If the eigen values
corresponding to the critical point are all negative, the critical
points are stable node, otherwise unstable. The eigen values for
the above critical points are obtained as in the following:\\

{\bf Table 3:} The eigen values corresponding to the critical
points ($x_{c},~y_{c},~z_{c}$).

\begin{center}
\begin{tabular}{|l|}
\hline\hline
~~NO:~~~~~~~~Value1~~~~~~~~~~~~~~~~~~~~~~~~~~~~~~~~~~~~~Value2~~~~~~~~~~~~~~~~~~~~~~~~~~~~~~~~~~~Value3
\\ \hline
\\\\
~~(i)~~~~~~~~~~$0$~~~~~~~~~~~~~~~$-\frac{3Ar_c^2+\sqrt{9A^2 r_c^4-3AQr_c^2\{-1+(-2r_c+\epsilon)^2\}\lambda}}{2Ar_c^2}$~~~~~~~~~~~~~$\frac{-3Ar_c^2+\sqrt{9A^2 r_c^4-3AQr_c^2\{-1+(-2r_c+\epsilon)^2\}\lambda}}{2Ar_c^2}$
\\\\
~~(ii)~~~~~~~~~$0$~~~~~~~~~~~~~~~$\frac{\sqrt{3\left[Ar_c^2\{6Ar_c^2+Q^2\{-1+(-2r_c+\epsilon)^2\}\}\right]}}{\sqrt{2}Ar_c^2}$~~~~~~~~~~~~~$-\frac{ \left(6Ar_c^2+\sqrt{6\left[Ar_c^2\{6Ar_c^2+Q^2\{-1+(-2r_c+\epsilon)^2\}\} \right]}\right)\lambda}{4 Ar_c^2}$
\\\\
~~(iii)~~~~~~~~~$0$~~~~~~~~~~~~~~~$-\frac{\sqrt{3\left[Ar_c^2\{6Ar_c^2+Q^2\{-1+(-2r_c+\epsilon)^2\}\}\right]}}{\sqrt{2}Ar_c^2}$~~~~~~~~~~~~~$\frac{ \left(-6Ar_c^2+\sqrt{6\left[Ar_c^2\{6Ar_c^2+Q^2\{-1+(-2r_c+\epsilon)^2\}\} \right]}\right)\lambda}{4 Ar_c^2}$
\\\\

\\\hline\hline
\end{tabular}
\end{center}

From Table 3, we see that one eigen value for all three critical
points is zero. Hence the dynamical system is unstable around all
critical points.

\subsection{Basic Equations in RS II Brane World}

Randall and Sundrum \cite{{Randall1},{Randall2}} elucidate the
higher dimensional scenario by introducing a bulk-brane model
dubbed as RS II brane model.  They proposed that we live in a four
dimensional world (called 3-brane, a domain wall) which is
embedded in a 5D space time (bulk). All matter fields are confined
in the brane and gravity can only propagate in the bulk. In RS II
Brane world the modified Einstein equations in flat universe are

\begin{eqnarray}
3H^2=\Lambda_4+\kappa_4^2 \rho+\frac{\kappa_4^2}{2 \lambda_1} \rho^2+\frac{6}{\lambda_1 \kappa_4^2 }{\cal U}\\
2\dot{H}+3H^2=\Lambda_4-\kappa_4^2 p-\frac{\kappa_4^2}{2\lambda_1}\rho p-\frac{\kappa_4^2}{2 \lambda_1}\rho^2-\frac{2}{\lambda_1 \kappa_4^2}{\cal U}
\end{eqnarray}

Here $\kappa_4$ and $\Lambda_4$ are respectively 4D gravitational
constant and effective 4D cosmological constant. The dark
radiation ${\cal U}$ satisfies the relation
\begin{eqnarray}
\dot{{\cal U}}+4H{\cal U}=0
\end{eqnarray}

\subsubsection{Dynamical system}

Here we introduce the new variables
\begin{eqnarray}
x=\frac{\dot{\phi}}{\sqrt{6}H},  ~~~~y=\frac{e^{-\lambda \phi/2}}{\sqrt{3}H} ~~~~~z=\frac{\rho}{2\lambda_1}
\end{eqnarray}

Also we introduce the fraction density function as $\Omega_\phi=\frac{\rho_\phi}{3 H^2}$ and $\Omega_m=\frac{\rho_m}{3 H^2}$ satisfying
\begin{eqnarray}
\Omega_\phi+\Omega_m+\Omega_{RS II}=1
\end{eqnarray}

where, $\Omega_{RS II}=1-\frac{1}{\kappa_4^2 (1+ z)}$ is the
density parameter due to the effect of RS II brane world with,
\begin{eqnarray}
\Omega_\phi=Ax^2+V_0 y^2,~~~~~w_\phi=\frac{Ax^2-V_0y^2}{Ax^2+V_0 y^2}~~~~~~~~~~~~~~\\
w_{eff}=\kappa_4^2 (1+z)\left[A(1-w_m)x^2-V_0(1+w_m)y^2+\frac{w_m}{\kappa_4^2 (1+z)}\right]
\end{eqnarray}

In absence of the cosmological constant and dark radiation
($\Lambda_4={\cal U}=0$) the above equations reduce to the
dynamical system of equations as follows
\begin{eqnarray}
\frac{dx}{dN}=-3 x+\frac{3}{2}\kappa_4^2 x \left[\left(A x^2+V_0 y^2\right) z
+2 A x^2 (1+z)+\{1+w_m+(2+w_m)z\} \left(-A x^2-V_0 y^2+\frac{1}{\kappa_4^2 (1+z)}\right)\right] \nonumber\\
-\frac{\sqrt{3}\left[-Q\left(A x^2+V_0 y^2\right)+\frac{Q}{\kappa_4^2 (1+z)}-V_0 \lambda y^2\right]}{\sqrt{2}A}~~~~~~~~~~~~~~~~~~~~~~~~~~~~~~~~~~
\end{eqnarray}

\begin{eqnarray}
\frac{dy}{dN}=\frac{y}{2}\left[6-\frac{3}{1+z}+3A\kappa_4^2
(1-w_m)x^2 (1+z)-3 \kappa_4^2V_0 y^2 (1+z)-3w_m\{-1+\kappa_4^2V_0
y^2(1+z)\}-\sqrt{6} \lambda x\right]
\end{eqnarray}

\begin{eqnarray}
\frac{dz}{dN}=-3\kappa_4^2 z(1+z)\left[A x^2(1-w_m)+(1+w_m)\left(\frac{1}{\kappa_4^2 (z+1)}-V_0y^2\right)\right]
\end{eqnarray}

\begin{eqnarray}
\frac{1}{H}\frac{dH}{dN}=-\frac{3}{2}\kappa_4^2\left[\left(A
x^2+V_0 y^2\right)z+2Ax^2 (1+z)+\{1+w_m+(2+w_m)z\}\left(-A x^2-V_0
y^2+\frac{1}{\kappa_4^2 (1+z)}\right)\right]
\end{eqnarray}

The new variables ($x,y,z$) has been drawn in figure 7 with
respect to $N=\ln a$ for RS II model. We see that $x,y,z$ are
shown to be positive throughout the evolution. The effective EoS
parameter $w_{eff}$ and the density parameters
$\Omega_{\phi},~\Omega_{m}$ are shown in figure 8. During
expansion, the $\Omega_{\phi}$ increases and $\Omega_{m}$
decreases which show the dark energy dominates at late times. Also
$w_{eff}$ decreases from some value $0.1$ to negative value $<-1$
which also show the dark energy dominated with phantom phase of the
universe.\\\\\\

\begin{figure}[!h]
\epsfxsize = 2.3 in \epsfysize = 1.7 in
\epsfbox{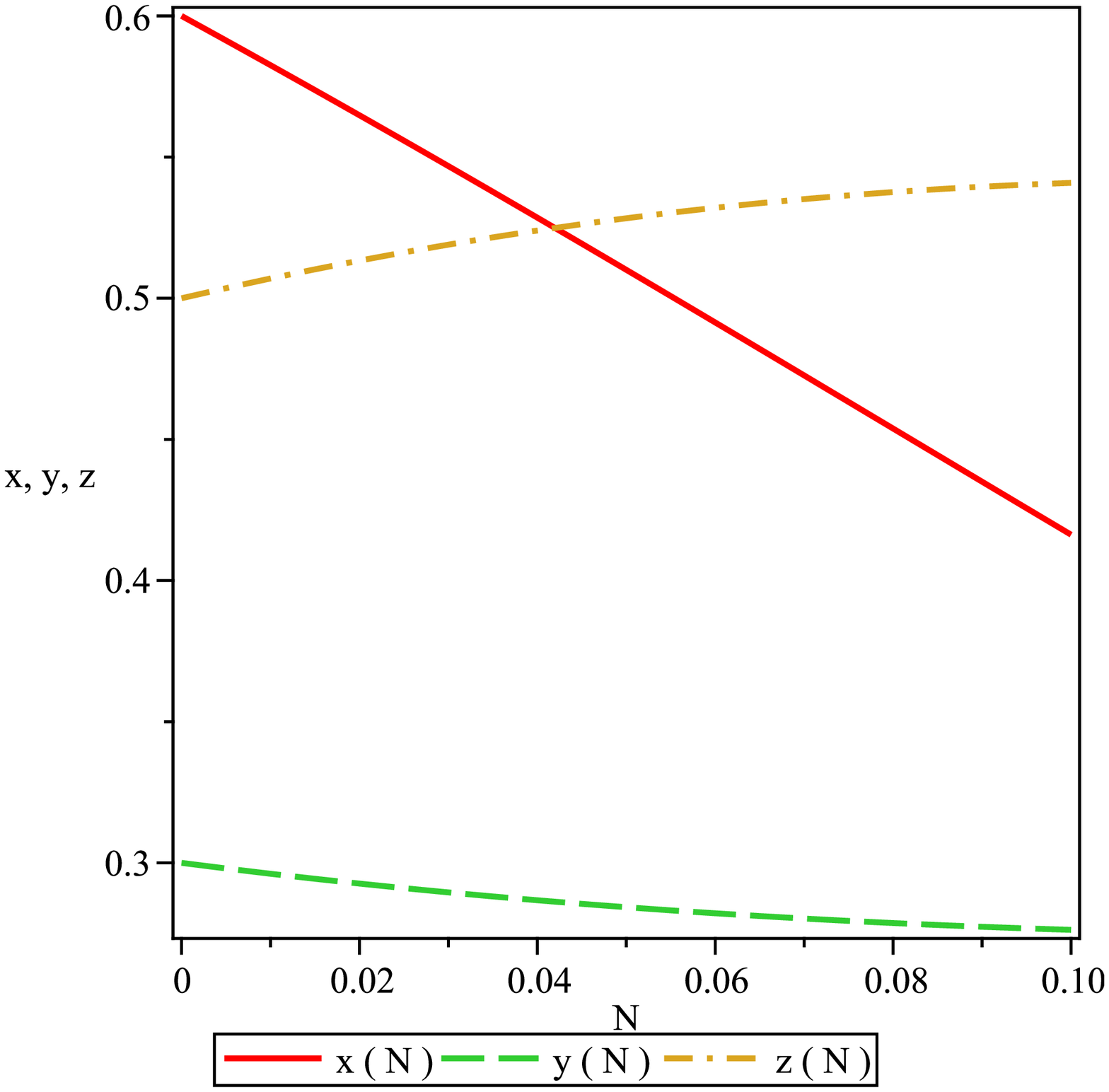}~~~\epsfxsize = 2.3 in \epsfysize =1.7 in
\epsfbox{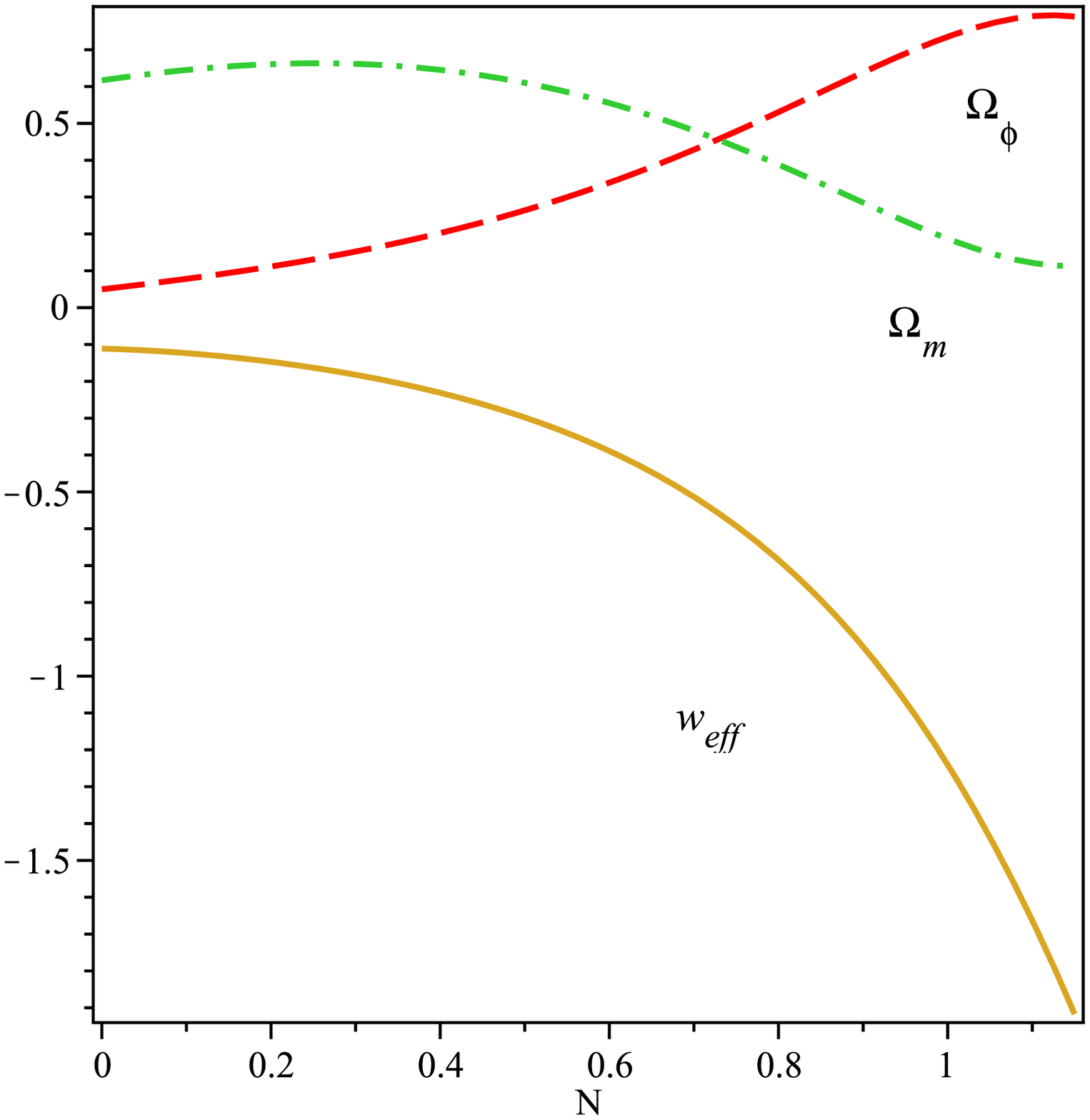}\\
~~FIG.7~~~~~~~~~~~~~~~~~~~~~~~~~~~~~~~~~~~~~~~~~~~~~~~~~~~~~~~~~~~~~~~~~~~~FIG.8\\
\caption{Evaluation of $x$, $y$, $z$ with respect to $N$ in RS II
Brane model taking $\gamma=0.6$, $Q=0.05$, $V_0=6.5$,
$\lambda=1.5$, $w_m=0.01$ and $\kappa_4=1$.} \caption{Evaluation
of $\Omega_\phi$, $\Omega_m$ and $w_{eff}$ with respect to $N$ in
RS II Brane model taking $\gamma=0.4$, $Q=0.2$, $V_0=6.5$,
$\lambda=1.5$, $w_m=0.01$ and $\kappa_4=1$.}
\end{figure}

\subsubsection{Critical points:}

The critical points can be obtained by setting $\frac{dx}{dN}=0$,
$\frac{dy}{dN}=0$ and $\frac{dz}{dN}=0$. The feasible critical
points ($x_{c},y_{c},z_{c}$) and the corresponding values of
$\Omega_{\phi}$ of our RS II model are given in the
following:\\

{\bf Table 4:} The critical points ($x_{c},~y_{c},~z_{c}$) and the
corresponding values of the density parameter $\Omega_{\phi}$.

\begin{center}
\begin{tabular}{|l|}
\hline\hline
~~No.~~$x_c$~~~~~~~~~~~~~~~~~~~~~~~~~~~$y_c$~~~~~~~~~~~~~~~~~~~~~~~~~~$z_c$~~~~~~~~~~~~~~~~~~~~~~~~~~~~~~~~~~~~~~~~~~~~~~~$\Omega_\phi$\\ \hline
\\\\
~~(i)~~$\frac{\sqrt{2} Q}{\sqrt{3} A \kappa_4^2 (w_m-1)}$~~~~~~~~~~~~~~~~~~~~0~~~~~~~~~~~~~~~~~~~~~~~~~~0~~~~~~~~~~~~~~~~~~~~~~~~~~~~~~~~~~~~~~~$\frac{2 Q^2}{3A \kappa_4^4 (w_m-1)^2}$\\\\
~~(ii)~$\frac{\sqrt{2}Q+\sqrt{2Q^2-3A \kappa_4^2 (w_m^2-1)}}{\sqrt{3}A \kappa_4^2 (w_m-1)}$~~~~~~0~~~~~~$-\frac{6A \kappa_4^2 (-1+w_m^2)+2Q (2Q+\sqrt{4Q^2-6A \kappa_4^2 (w_m^2-1)})}{3A \kappa_4^2 (w_m^2-1)}$~~~~~$\frac{\left(\sqrt{2}Q+\sqrt{2Q^2-3A \kappa_4^2 (w_m^2-1)}\right)^2}{3A \kappa_4^4 (w_m-1)^2}$\\\\
~~(iii)~$\frac{\sqrt{6}Q-\sqrt{6Q^2-9A \kappa_4^2 (w_m^2-1)}}{3A \kappa_4^2 (w_m-1)}$~~~~~~0~~~~~~$\frac{-6A \kappa_4^2 (-1+w_m^2)+2Q (2Q+\sqrt{4Q^2-6A \kappa_4^2 (w_m^2-1)})}{3A \kappa_4^2 (w_m^2-1)}$~~~~~~~~~$\frac{\left(-\sqrt{2}Q+\sqrt{2Q^2-3A \kappa_4^2 (w_m^2-1)}\right)^2}{3A \kappa_4^4 (w_m-1)^2}$\\\\
\\\hline\hline
\end{tabular}
\end{center}

From the Table 4, we see that the components of $y_{c}$ are equal
to zero for above three critical points. The value of
$\Omega_{\phi}>1$ or $<1$ for the critical points given in (i) to
(iii), depends on the values of the $w_{m}$, $A$ and the
interaction term $Q$.\\

\subsubsection{Stability of the model:}

Now the stability around the critical points can by determined by
the sign of the corresponding eigen values. If the eigen values
corresponding to the critical point are all negative, the critical
points are stable node, otherwise unstable. The eigen values for
the first critical point (i) are obtained as in the following.
The eigen values of critical points (ii) and (iii) are very difficult
to obtain, so we have not considered that critical points
here.\\\\\\\\

{\bf Table 5:} The eigen values corresponding to the first
critical point ($x_{c},~y_{c},~z_{c}$).

\begin{center}
\begin{tabular}{|l|}
\hline\hline
~~NO:~~~~~~~~Value1~~~~~~~~~~~~~~~~~~~~~~~~~~~~~~~~~~~~~Value2~~~~~~~~~~~~~~~~~~~~~~~~~~~~~~~~~~~Value3
\\ \hline
\\\\
~~(i)~~~~~~~~~~$\frac{2 Q^2}{A \kappa_4^2 (w_m-1)}-3 (1+w_m)$~~~~~~~~~~~~~~~$-\frac{Q^2}{A \kappa_4^2 (w_m-1)}-\frac{3}{2} (1-w_m)$~~~~~~~~~~~~~$\frac{3A \kappa_4^2(-1+w_m^2)-2Q (Q+\lambda)}{2 A \kappa_4^2 (-1+w_m)}$
\\\\
\\\hline\hline
\end{tabular}
\end{center}

At the critical point (i), the three eigen values cannot be
negative simultaneously, since $Q$ and $w_{m}$ are small
quantities. So the dynamical system is unstable in this case. At
the critical points (ii) and (iii) we can not study the stability
analysis.

\section{Discussions}

In this work, we have studied homogeneous isotropic FRW model
having dynamical dark energy DBI-essence with scalar field in
presence of perfect fluid having barotropic equation of state
(i.e., $p_{m}=w_{m}\rho_{m}$). The existence of cosmological
scaling solutions restricts the Lagrangian of the scalar field
$\phi$. The stable attractor solution can be found only for
$\gamma =$ constant. We have chosen the potential function for
DBI-essence as $V(\phi)=V_{0}e^{\lambda\phi}$. Choosing $p=X g(X
e^{\lambda \phi})$, where $X=-g^{\mu\nu}
\partial_\mu \phi
\partial_\nu \phi /2$ with $g$ is any function of $X e^{\lambda
\phi}$ and defining some suitable transformations, we have
constructed the dynamical system in different gravity theories
like (i) Loop Quantum Cosmology (LQC), (ii) DGP Brane World and
(iii) RS-II Brane World. For all gravity models, $\Omega_{m}$
gradually decreases to a small positive value and $\Omega_{\phi}$
gradually increases to a value near about 1. That means, DBI dark
energy dominates over dark matter in late times. Also from the
figures of $w_{eff}$, we see that $w_{eff}$ keeps negative sign in
late times. For LQC model, $w_{eff}$ lies between $-0.5$ and $-1$,
which is the dark energy dominated phase. For DGP model, $w_{eff}$
lies between $-0.4$ and $-1$. So for LQC and DGP models of the
universe, the DBI dark energy valid only for quintessence era,
they can not generate phantom era. Also in RS II model,
$w_{eff}<-0.1$ and decreases to $-1$ upto certain stage of time
and after that stage $w_{eff}$ becomes less than $-1$. So in RS II
model, the DBI dark energy valid for quintessence era and phantom
era in late times. We have found some critical points and
investigated the stability of this dynamical system around the
critical points for three gravity models and investigated the
scalar field dominated attractor solution in support of
accelerated universe. Gumjudpai et al \cite{Naskar} in their work
considered the dynamical system analysis for phantom field,
tachyonic field and dilaton models of dark energy.  They analyzed
different dark energy model of scalar field coupled with
barotropic perfect fluid and depicts that scaling solution is
stable if the state parameter $w_\phi>-1$ and the scalar field
dominated solution becomes unstable. The fixed points are always
classically stable for a phantom field, implying that the universe
is eventually dominated by the energy density of a scalar field if
phantom is responsible for dark energy. Therefore in this case the
final attractor is either a scaling solution with constant
$\Omega_{\phi}$ satisfying $0 <\Omega_{\phi} < 1$ or a
scalar-field dominant solution with $\Omega_{\phi} = 1$. For our
LQC model, four critical points have been found, in which only two
critical points may be stable node while all other two critical
points are unstable. For DGP model, three critical points have
been found but they are all unstable. Also for RS II model, the
calculated critical point is also unstable. An attractor scaling
is established by Martin and Yamaguchi \cite{Martin} after
considering a dark energy model with DBI field. Recent times, the
model of interacting dark energy has been explored in the
framework of loop quantum cosmology (LQC). On that framework an
interacting MCG with dark matter has been studied by constructing
a dynamical system and depicts a scaling attractor solution which
resolve the cosmic coincidence problem in modern Cosmology
\cite{jamil}. A dynamical system is explored in DGP, RSII Brane
world separately with suitable interacting dark energy coupled
with dark matter model \cite{Rudra} and investigated that the
universe in both scenarios follow the power law form of expansion
around the critical point. So in conclusion, DBI-essence plays an
important role of dark energy for FRW model of the universe in
loop quantum cosmology, which drives the acceleration of the universe. \\\\

{\bf Acknowledgement:}\\

The authors are thankful to IUCAA, Pune, India for warm
hospitality where part of the work was carried out. One of
the authors (JB) is thankful to CSIR, Govt of India for
providing Junior Research Fellowship.\\

\end{document}